\documentclass{article}

\usepackage{arxiv}

\usepackage[utf8]{inputenc} 
\usepackage[T1]{fontenc}    
\usepackage{hyperref}       
\usepackage{url}            
\usepackage{booktabs}       
\usepackage{amsfonts}       
\usepackage{nicefrac}       
\usepackage{microtype}      
\usepackage{lipsum}		
\usepackage{graphicx}
\usepackage{doi}
\usepackage{amsmath}
\usepackage{authblk}
\usepackage[title]{appendix}
\usepackage[sort&compress,numbers]{natbib}

\title{On the closure of curvature in 2D flamelet theory}
\author[ ~,1]{Hernan Olguin\thanks{Corresponding author\\ Email address: hernan.olguin@usm.cl}}
\author[2]{Pascale Domingo}
\author[2]{Luc Vervisch}
\author[3]{Christian Hasse}
\author[3]{Arne Scholtissek}

\affil[1]{Department of Mechanical Engineering, Universidad Técnica Federico Santa María,\linebreak Avenida España 1680, Valparaíso, Chile\linebreak}
\affil[2]{CORIA CNRS,  Normandie Universit\'e, INSA de Rouen, Technop\^ole du Madrillet, BP 8,\linebreak Saint-\'Etienne-du-Rouvray 76801, France\linebreak}
\affil[3]{Institute for Simulation of reactive Thermo-Fluid Systems, TU Darmstadt,\linebreak Otto-Berndt-Stra{\ss}e 2, 64287 Darmstadt, Germany}

\date{}

\begin{document}
\maketitle

\begin{abstract}
So far, flamelet theory has treated curvature as an independent parameter requiring specific means for closure. In this work, it is shown how the adoption of a two-dimensional orthogonal composition space allows obtaining formal mathematical relations between the flame curvatures and the gradients of the conditioning scalars (also called flamelet coordinates). With these, both curvatures become a flame response to the underlying flow, which conveniently allows removing them from the corresponding set of flamelet equations. While the demonstration is performed in the context of partially premixed flames, the approach is general and applicable to any orthogonal coordinate system. 
\end{abstract}

\vspace{15mm}

\section{Introduction}\label{introduction}

Orthogonal coordinate systems have been recently proposed as an ideal framework for the derivation of flamelet equations for partially premixed flames:  Their adoption avoids the need of closure means for the cross scalar dissipation rate and leads to a formulation allowing for a direct recovery of the asymptotic limits of non-premixed and premixed combustion~\citep{Scholtissek20,Olguin232}. In this context, it has been shown how, after introducing the mixture fraction, $Z$, as main coordinate, a modified reaction progress variable, $\varphi$, can be defined in such a way that it satisfies the orthogonality condition, $\nabla Z \cdot \nabla \varphi = 0~$\cite{Olguin232}. Based on the ($Z, \varphi$)-space, flamelet equations for the chemical species mass fraction, temperature, and both conditioning scalar gradients, $g_Z = \lvert \nabla Z \rvert$ and $g_{\varphi} = \lvert \nabla \varphi \rvert$, have been obtained in terms of four parameters: Two strain rates and two curvatures.

Classical combustion theory naturally contains strain and curvature components, such that the latter is often considered a parameter, rather than a flame response (see for example~\cite{Matalon83,Chung84,DeGoey97}). Similarly, flamelet theory has so far treated flame curvatures as independent parameters~\cite{Kortschik05,Xu13,Xuan14,Scholtissek15}. However, it will be shown now that the adoption of an orthogonal coordinate system has an additional yet unexplored advantage, namely the fact that it allows connecting the curvatures with the derivatives of the scalar gradients, $g_Z$ and $g_{\varphi}$. With this, the two strain rates are the only parameters remaining in the formulation proposed in~\cite{Olguin232}, while curvature becomes a flame response to the underlying flow. 

\section{The relation between the curvatures and the conditioning scalar gradients}
\label{section2}

We start the derivation considering a flamelet-like transformation from a two-dimensional physical space, ($t, x, y$) into a corresponding composition space, ($\tau, Z, \varphi$). Here, $\tau$ is a time-like variable and the mixture fraction, $Z$, and the modified reaction progress variable, $\varphi$, are formally defined through their respective governing equations 
\begin{equation}
    \frac{\partial Z}{\partial t} + \mathbf{u} \cdot \nabla Z = \frac{1}{\rho} \nabla \cdot \left( \rho D \nabla Z \right)
    \label{mixturefraction}
\end{equation}
and
\begin{equation}
     \frac{\partial \varphi}{\partial t} + \mathbf{u} \cdot \nabla \varphi = \frac{1}{\rho} \nabla \cdot \left( \rho D \nabla \varphi \right) + \frac{\dot{\omega}_{\varphi}}{\rho},
     \label{progressvariable}
\end{equation}
where $\mathbf{u}$ is the flow velocity, $\rho$ denotes the gas density and $D$ corresponds to a diffusion coefficient. Moreover, the source term in Eq.~(\ref{progressvariable}) is defined as
\begin{equation}
    \dot{\omega}_{\varphi} = \dot{\omega}_c + \rho D \lvert \nabla Z \rvert^2 \frac{\partial^2 Y_c}{\partial Z^2},
    \label{sourceterm}
\end{equation}
where the conventional reaction progress variable, $Y_c$, is defined as a suitable combination of (product) species mass fractions. With Eq.~(\ref{sourceterm}), it is ensured that the orthogonality condition, $\nabla Z \cdot \nabla \varphi = 0$, is satisfied (see formal derivation in~\cite{Olguin232}). 

Based on $Z$ and $\varphi$, two unit vectors can be introduced now as
\begin{equation}
    \mathbf{n}_Z = \frac{\nabla Z}{\lvert \nabla Z \rvert}~\text{      }~\text{   and     }~\text{       }~\mathbf{n}_{\varphi} = \frac{\nabla \varphi}{\lvert \nabla \varphi \rvert},
\end{equation}
which allows defining the two associated curvatures
\begin{equation}
    \kappa_Z = - \nabla \cdot \mathbf{n}_Z~\text{      }~\text{   and     }~\text{       }~\kappa_{\varphi} = - \nabla \cdot \mathbf{n}_{\varphi}.
    \label{kappa_def}
\end{equation}
For simplicity, in the rest of this section we will focus on $\kappa_{\varphi}$ and its relation with $g_Z = \lvert \nabla Z \rvert$, but the analysis can be replicated to study the relation between $\kappa_{Z}$ and $g_{\varphi} = \lvert \nabla \varphi \rvert$.

Now, the orthogonality between $\mathbf{n}_Z$ and $\mathbf{n}_{\varphi}$ allows relating the components of these two vectors. For example, expressing them in the following generic form
\begin{equation}
\mathbf{n}_{Z} = n_x \mathbf{e}_{x} + n_y \mathbf{e}_{y}~\text{      }~\text{   and     }~\text{       }~\mathbf{n}_{\varphi} = m_x \mathbf{e}_{x} + m_y \mathbf{e}_{y},
\label{nZ}
\end{equation}
where $\mathbf{e}_x = (1,0)^T$ and $\mathbf{e}_y = (0,1)^T$, it is clear that the required orthogonality can be satisfied setting $m_x = n_y$ and $m_y = -n_x$, since 
\begin{equation}
(\mathbf{n}_{Z} \cdot \mathbf{n}_{\varphi}) = n_x n_y - n_y n_x = 0.
\end{equation}
With this,  $\kappa_{\varphi}$ can be rewritten as
\begin{equation}
  \kappa_{\varphi} = - \frac{\partial n_y}{\partial x} + \frac{\partial n_x}{\partial y},
  \label{kappa1}
\end{equation}
where the derivatives at the RHS correspond to different components of the curvature tensor, $\nabla \mathbf{n}_Z$. These can be further worked out in terms of $Z$ by means of the following identity~\cite{Dopazo07} (see Appendix~\ref{appendix} for a detailed derivation)
\begin{equation}
    \frac{\partial n_i}{\partial x_j}= \frac{1}{g_Z} \left[ \frac{\partial^2 Z}{\partial x_i \partial x_j }  - n_i n_k \frac{\partial^2 Z}{\partial x_j x_k} \right],
    \label{relation-appendix}
\end{equation}
which yields
\begin{equation}
    \frac{\partial n_y}{\partial x} = \frac{1}{g_{Z}} \left[ \frac{\partial^2 Z}{\partial x \partial y} - n_x n_y \frac{\partial^2 Z}{\partial x^2} - n_y^2 \frac{\partial^2 Z}{\partial x \partial y}\right]
\end{equation}
and
\begin{equation}
    \frac{\partial n_x}{\partial y} = \frac{1}{g_{Z}} \left[ \frac{\partial^2 Z}{\partial x \partial y} - n_x^2 \frac{\partial^2 Z}{\partial x \partial y} - n_x n_y \frac{\partial^2 Z}{\partial y^2}\right],
\end{equation}
respectively. Replacing back in Eq.~(\ref{kappa1}), we obtain
\begin{equation}
    \kappa_{\varphi} = \frac{1}{g_Z} \left[n_x n_y \left( \frac{\partial^2 Z}{\partial x^2} - \frac{\partial^2 Z}{\partial y^2}\right) + \left( n_y^2 - n_x^2 \right) \frac{\partial^2 Z}{\partial x \partial y}\right],
    \label{kappa2}
\end{equation}
where the term in brackets at the RHS of this equation corresponds to $\partial g_Z / \partial n_{\varphi} = g_{\varphi} \partial g_Z / \partial \varphi$, as it will be shown next. 

Based on the definition of the directional derivative, we can write
\begin{equation}
    \frac{\partial g_Z}{\partial n_{\varphi}} = \mathbf{n}_{\varphi} \cdot \nabla g_Z = n_y \frac{\partial g_Z}{\partial x} - n_x \frac{\partial g_Z}{\partial y},
    \label{directionalderivative}
\end{equation}
where the derivatives at the RHS can be further worked out by means of the following mathematical identity (see Appendix~\ref{appendix} for a detailed derivation)
\begin{equation}
    \frac{\partial g_{Z}}{\partial x_j} = n_i \frac{\partial^2 Z}{\partial x_i \partial x_j},
    \label{identitymain2}
\end{equation}
which yields
\begin{equation}
    \frac{\partial g_Z}{\partial x} = n_x \frac{\partial^2 Z}{\partial x^2} + n_y \frac{\partial^2 Z}{\partial x \partial y}
\end{equation}
and
\begin{equation}
    \frac{\partial g_Z}{\partial y} = n_x \frac{\partial^2 Z}{\partial x \partial y} + n_y \frac{\partial^2 Z}{\partial y^2},
\end{equation}
respectively. Replacing in Eq.~(\ref{directionalderivative}), we have
\begin{equation}
    \frac{\partial g_{Z}}{\partial n_{\varphi}} = n_x n_y \left( \frac{\partial^2 Z}{\partial x^2} - \frac{\partial^2 Z}{\partial y^2} \right) + \left(n_y^2 - n_x^2 \right) \frac{\partial^2 Z}{\partial x \partial y},
\end{equation}
which can be inserted in Eq.~(\ref{kappa2}) to obtain
\begin{equation}
    \kappa_{\varphi} = \frac{g_{\varphi}}{g_{Z}} \frac{\partial g_{Z}}{\partial \varphi},
    \label{kappa3}
\end{equation}
which, together with its equivalent expression for $\kappa_{Z}$, will allow removing both curvatures as parameters in the corresponding flamelet equations, as it will be shown in Section~\ref{2D-flamelets}. 

At this point, two important aspects must be highlighted. First, the derivation shown in this section is not the only possible path to obtain Eq.~(\ref{kappa3}) (see for example Appendix \ref{alternative} for an alternative derivation). Secondly, it is interesting that Eq.~(\ref{kappa3}) can be recast as one of the terms appearing in the multi-dimensional flamelet equation obtained in~\cite{wi-85} (see Eq.~(92), page~77). This yields
\begin{equation}
\kappa_{\varphi} = \frac{g_\varphi}{g_Z} \frac{\partial g_Z}{\partial\varphi}=g_\varphi \frac{\partial \ln g_Z}{\partial\varphi}= \nabla_T \ln g_Z \:,
\end{equation}
where $\nabla_T (\cdot)$ corresponds to Williams' notation for the above-defined directional derivative $\partial (\cdot) / \partial n_{\varphi}$. Thus, the current approach also provides new physical insights into the classical flamelet formulation presented in~\cite{wi-85}.

\section{Numerical validation for a triple flame}

\begin{figure}[b]
    \centering
    \vspace{-8pt}
    \includegraphics[scale=0.5]{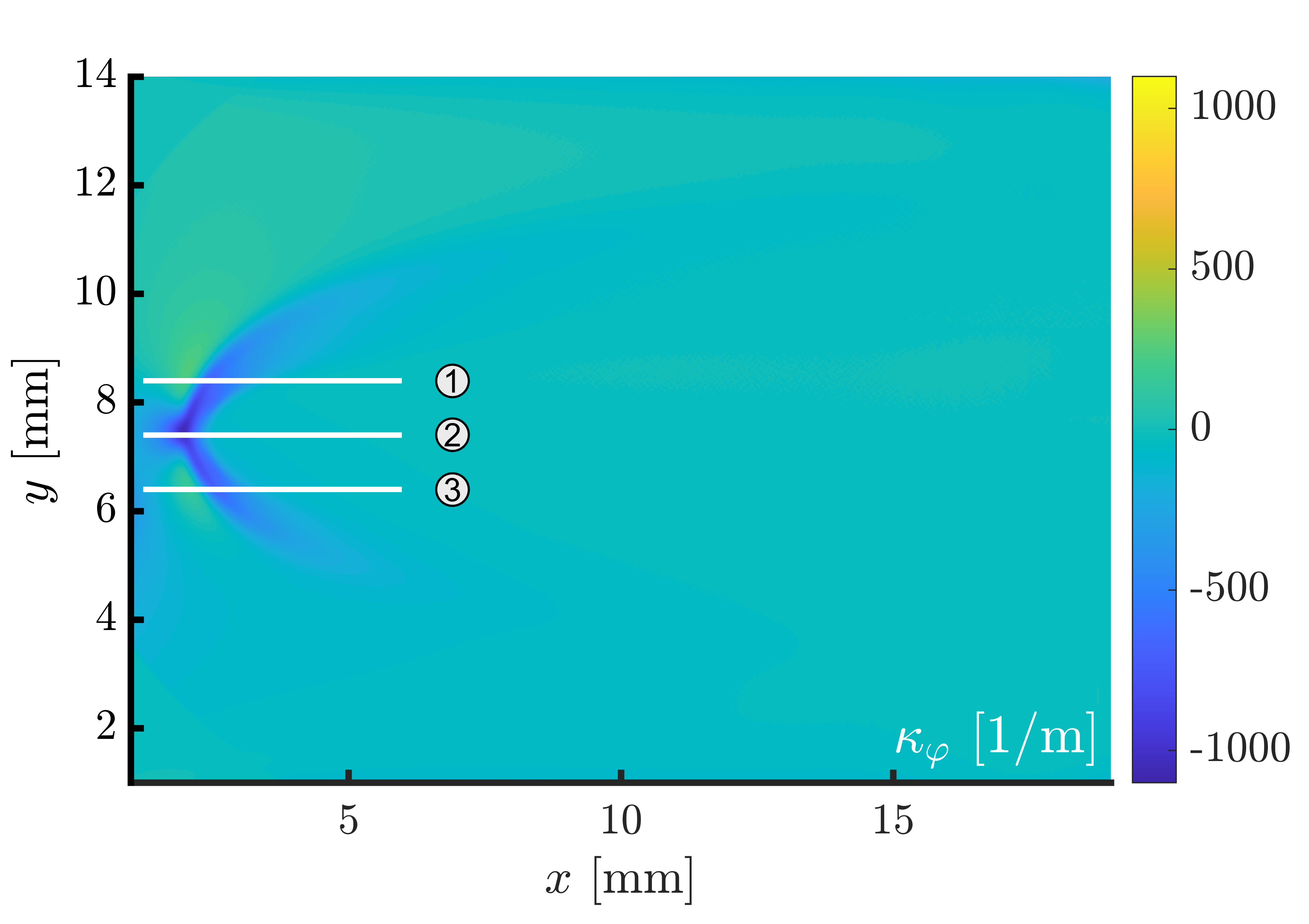}
    \vspace{-8pt}
    \caption{$\kappa_{\varphi}$ scalar field evaluated for the methane-air triple flame.}
    \label{fig:curv_contour}
\end{figure}

For the verification of Eq.~(\ref{kappa3}), we analyze a methane-air triple flame previously studied in~\cite{Scholtissek20}. This flame is established by the consideration of an inflow of premixed fresh gases at atmospheric conditions (300~K and 1~atm) with a mixture stratification in the cross-flow direction. The minimum and maximum mixture fractions at the inlet are 0 and 0.42, respectively, while the imposed mixture fraction gradient is 50~m$^{-1}$. For more details on this flame, the reader is referred to~\cite{Scholtissek20}.

\begin{figure}[t]
    \centering
    \includegraphics[scale=0.5]{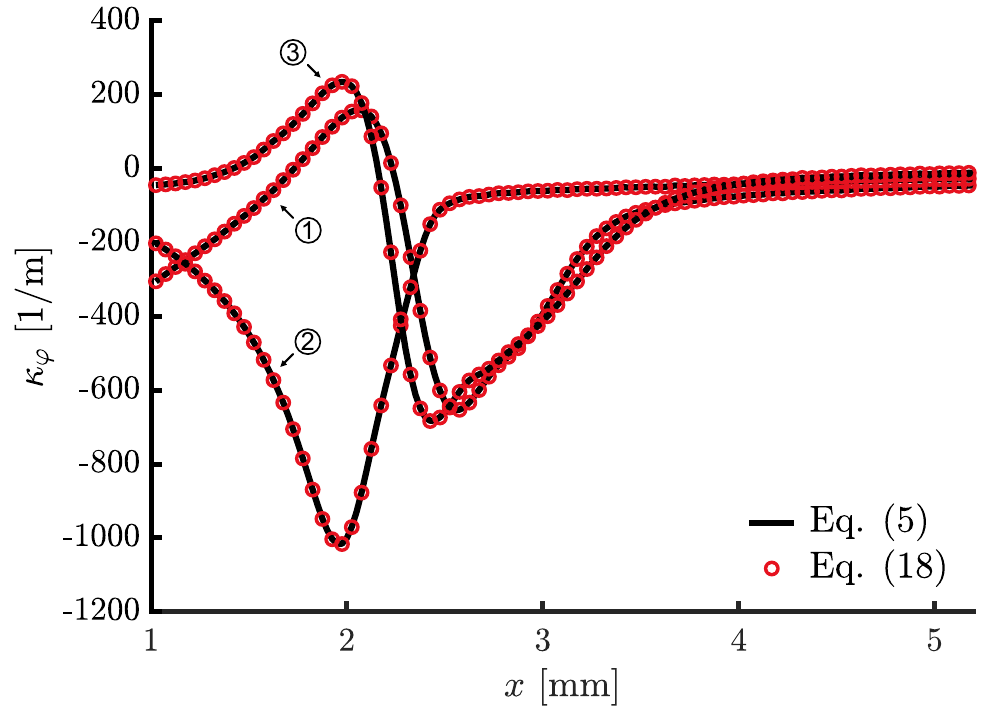}
    \caption{Profiles of $\kappa_{\varphi}$ along the slices shown in Fig.~\ref{fig:curv_contour}.}
    \vspace{-8pt}
    \label{fig:curv_profiles}
\end{figure}

Figure~\ref{fig:curv_contour} displays the $\kappa_{\varphi}$ field associated with the chosen flame (obtained by direct evaluation of Eq.~(\ref{kappa_def})), where three different horizontal slices are identified as representative regions for the aimed validation. In Fig.~\ref{fig:curv_profiles}, the corresponding comparison between Eqs.~(\ref{kappa_def}) and (\ref{kappa3}) along the selected slices is shown, where a perfect match is observed. In this way, the validity of the analysis presented in Section~\ref{section2} is numerically confirmed. 

\section{Two-dimensional flamelet equations with $a_Z$ and $a_{\varphi}$ as only parameters}
\label{2D-flamelets}

Making use of the obtained relations between the curvatures and the gradients of the conditioning scalars, the two-dimensional flamelet equations derived in~\cite{Olguin232} for the chemical species mass fractions, $Y_k$, and the temperature, $T$, can be rewritten as
\begin{align}
\rho \frac{\partial Y_k}{\partial \tau}  & = -  \frac{\partial Y_k}{\partial \varphi}  \dot{\omega}_{\varphi}  + \rho D g_Z^2  \frac{\partial^2 Y_k}{\partial Z^2} + \rho D g_{\varphi}^2 \frac{\partial^2 Y_k}{\partial \varphi^2} + \dot{\omega}_k
\label{e15}
\end{align}
and
\begin{align}\label{e16}
\rho \frac{\partial T}{\partial \tau} & = - \frac{\partial T}{\partial \varphi}  \dot{\omega}_{\varphi} + \rho D g_Z^2 \frac{\partial^2 T}{\partial Z^2} +  \rho D g_{\varphi}^2 \frac{\partial^2 T}{\partial \varphi^2} + \dot{\omega}_T \nonumber \\ & + \frac{\rho D}{ c_p}  \left[  g_Z^2 \frac{\partial T}{\partial Z} \frac{\partial c_p}{\partial Z} + g_{\varphi}^2  \frac{\partial T}{\partial \varphi} \frac{\partial c_p}{\partial \varphi}   \right]  \\ &+  \sum_{k=1}^{N}  \frac{ c_{p,k}}{ c_p}  \left[ \rho D g_Z^2  \frac{\partial Y_k}{\partial Z} \frac{\partial T}{\partial Z} + \rho D g_{\varphi}^2  \frac{\partial Y_k}{\partial \varphi} \frac{\partial T}{\partial \varphi} \right], \nonumber
\end{align} 
respectively. Similarly, the corresponding equations for $g_Z$ and $g_{\varphi}$ become
\begin{align}
    \frac{\partial g_Z}{\partial \tau} & = - \left[\frac{g_\varphi}{\rho}\frac{\partial}{\partial\varphi}\left(\rho D g_\varphi\right) - D \frac{g_{\varphi}^2}{g_{Z}} \frac{\partial g_{Z}}{\partial \varphi} + \frac{\dot{\omega}_{\varphi}}{\rho} \right] \frac{\partial g_Z}{\partial \varphi} \nonumber \\ & + \frac{g_Z^2}{\rho}\frac{\partial^2}{\partial Z^2}\left(\rho D g_Z\right) - \frac{g_Z^2}{\rho^2}\frac{\partial \rho}{\partial Z}\frac{\partial}{\partial Z}\left(\rho D g_Z\right) \nonumber \\ & - g_Z^2\frac{\partial}{\partial Z}\left(D \frac{g_{Z}}{g_{\varphi}} \frac{\partial g_{\varphi}}{\partial Z} \right) + g_Z a_Z 
    \label{egZ}
\end{align}
and
\begin{align}
    \frac{\partial g_\varphi}{\partial \tau} & = -  \left[ \frac{g_Z}{\rho}\frac{\partial}{\partial Z}\left(\rho D g_Z\right) - D \frac{g_{Z}^2}{g_{\varphi}} \frac{\partial g_{\varphi}}{\partial Z} \right] \frac{\partial g_{\varphi}}{\partial Z} \nonumber \\ & + \frac{g_\varphi^2}{\rho}\frac{\partial^2}{\partial \varphi^2}\left(\rho D g_\varphi\right) - \frac{g_\varphi^2}{\rho^2}\frac{\partial \rho}{\partial \varphi}\frac{\partial}{\partial \varphi}\left(\rho D g_\varphi\right) \nonumber \\ & - g_\varphi^2 \frac{\partial}{\partial \varphi} \left( D \frac{g_{\varphi}}{g_{Z}} \frac{\partial g_{Z}}{\partial \varphi} \right) + g_\varphi^2 \frac{\partial}{\partial \varphi} \left( \frac{\dot{\omega}_\varphi}{\rho g_\varphi} \right) + g_\varphi a_\varphi.
    \label{egvarphi}
\end{align}
As highlighted before, in these equations the only parameters to be imposed are the strain rates $a_Z = - \mathbf{n}_Z \cdot \nabla \mathbf{u} \cdot \mathbf{n}_Z$ and $a_{\varphi} = - \mathbf{n}_{\varphi} \cdot \nabla \mathbf{u} \cdot \mathbf{n}_{\varphi}$, while both curvatures can be now calculated as a flame response.

\section{Conclusions}

In this work, the recently proposed ($Z, \varphi$) flamelet space has been used to illustrate a so far unnoticed feature common to any orthogonal composition space coordinate system. More specifically, it has been shown how the space orthogonality allows deriving explicit relations between the curvatures, $\kappa_Z$ and $\kappa_{\varphi}$, and the gradients of the conditioning scalars, $g_Z$ and $g_{\varphi}$. Making use of these relations, both curvatures can be conveniently removed from the corresponding set of two-dimensional orthogonal flamelet equations derived in~\cite{Olguin232}. With this, the only parameters remaining in the formulation are the two strain rates, $a_Z$ and $a_{\varphi}$, solving in this way all problems associated with the closure of curvature in these equations. 

\appendix 
\numberwithin{equation}{section}
\begin{appendices}
\renewcommand{\appendixname}{Appendix}
\section{Mathematical identities}
\label{appendix}
\renewcommand{\appendixname}{}
The different components of the curvature tensor, $\nabla \mathbf{n}_Z$, can be expressed as
\begin{align}
    \frac{\partial n_i}{\partial x_j} & = \frac{\partial}{\partial x_j} \left( \frac{1}{g_{Z}} \frac{\partial Z}{\partial x_i}\right) \nonumber \\ & = \frac{1}{g_{Z}} \frac{\partial^2 Z}{\partial x_i \partial x_j} - \frac{1}{g_{Z}^2} \frac{\partial Z}{ \partial x_i}\frac{\partial g_{Z}}{\partial x_j},
    \label{rel1}
\end{align}
where the second terms at the RHS can be further worked out as
\begin{align}
   - \frac{1}{g_{Z}^2} \frac{\partial Z}{ \partial x_i}\frac{\partial g_{Z}}{\partial x_j} & = - \frac{1}{g_{Z}} n_i \frac{\partial}{\partial x_j} \left( n_k \frac{\partial Z}{\partial x_k}\right) \nonumber \\ & = - \frac{1}{g_{Z}} n_i n_k \left(  \frac{\partial^2 Z}{\partial x_j \partial x_k}\right) - n_i \underbrace{n_k \frac{\partial n_k}{\partial x_j}}_{= 0}
   \label{rel2}
\end{align}
Replacing Eq.~(\ref{rel2}) into (\ref{rel1}), we obtain
\begin{equation}
    \frac{\partial n_i}{\partial x_j}= \frac{1}{g_Z} \left[ \frac{\partial^2 Z}{\partial x_i \partial x_j }  - n_i n_k \frac{\partial^2 Z}{\partial x_j x_k} \right],
    \label{identity1}
\end{equation}
which corresponds to Eq.~(\ref{relation-appendix}).

Additionally, the derivatives of $g_Z$ in physical space can be rewritten as
\begin{equation}
    \frac{\partial g_{Z}}{\partial x_j} = \frac{\partial}{\partial x_j} \left(n_i \frac{\partial Z}{\partial x_i} \right),
    \label{identity2222}
\end{equation}
where the RHS can be split as 
\begin{equation}
     \frac{\partial}{\partial x_j} \left(n_i \frac{\partial Z}{\partial x_i} \right) = n_i \frac{\partial^2 Z}{\partial x_i \partial x_j} + g_Z \underbrace{n_i \frac{\partial n_i}{\partial x_j}}_{=0}. 
\end{equation}
Replacing back in Eq.~(\ref{identity2222}), we obtain
\begin{equation}
    \frac{\partial g_{Z}}{\partial x_j} = n_i \frac{\partial^2 Z}{\partial x_i \partial x_j},
\end{equation}
which corresponds to Eq.~(\ref{identitymain2}).

\renewcommand{\appendixname}{Appendix}
\section{Alternative derivation of Eq.~(\ref{kappa3})}
\label{alternative}
\renewcommand{\appendixname}{}

In the orthogonal space built by the vectors $\mathbf{n}_Z$ and $\mathbf{n}_{\varphi}$, the curvature of the $\varphi$-isosurfaces can be expressed as
\begin{align}
    \kappa_{\varphi} & = - \nabla \cdot \mathbf{n}_{\varphi} \nonumber \\ & = - \frac{\partial \mathbf{n}_{\varphi} }{\partial n_Z} \cdot \mathbf{n}_Z - \underbrace{\frac{\partial \mathbf{n}_{\varphi}}{\partial n_{\varphi}} \cdot \mathbf{n}_{\varphi}}_{= 0}.
    \label{kappa_app}
\end{align}
Making use of the space orthogonality, we obtain the following identity
\begin{equation}
    \frac{\partial ( \nabla Z \cdot \mathbf{n}_{\varphi} )}{\partial n_Z} =  \nabla Z \cdot \frac{\partial \mathbf{n}_{\varphi}}{\partial n_Z} + \mathbf{n}_{\varphi} \cdot \frac{\partial \nabla Z}{\partial n_Z} = 0,
\end{equation}
which can be used to obtain Eq.~(\ref{kappa3}) from Eq.~(\ref{kappa_app}) as
\begin{align}
    \kappa_{\varphi} & = \frac{1}{g_Z} \mathbf{n}_{\varphi} \cdot \frac{\partial \nabla Z}{\partial n_Z} \nonumber \\ & = \frac{1}{g_Z} \mathbf{n}_{\varphi} \cdot \nabla \left( \frac{\partial Z}{\partial n_Z} \right) \nonumber \\ & = \frac{g_{\varphi}}{g_Z} \frac{\partial g_Z}{\partial\varphi},
\end{align}
where use has been made of the flamelet transformation $\mathbf{n}_{\varphi} \cdot \nabla (\cdot) = \partial (\cdot) / \partial n_{\varphi} = g_{\varphi} \partial (\cdot) / \partial \varphi$.

\end{appendices}

\bibliography{main}










\end{document}